\documentclass{article}
\usepackage{subeqn}
\newcommand{\beeq}{\begin{equation}}
\newcommand{\eneq}{\end{equation}}
\newcommand{\beqn}{\begin{eqnarray}}
\newcommand{\eeqn}{\end{eqnarray}}
\def\qv{{\bf{q}}}
\def\SBZ{\sum_{\qv \in BZ}}
\def\gqv{\gamma_{\qv}}
\def\hc{\hat{c}}
\def\hcd{\hc^\dagger}
\def\hh{\hat{H}}
\def\om{\Omega_{m}}
\def\qom{(\qv,\om)}
\def\mv{\mbox{\boldmath{$\cal{M}$}}}
\def\mesm{\cal{D}\mu_{\mbox{\boldmath{$\cal{\scriptscriptstyle M}$}}}}
\def\ra{\rangle}
\def\la{\langle}
\def\dmv{\delta \mv}
\def\bS{{\bf S}}
\def\hbS{\hat{\bS}}
\def\bsigma{\mbox{\boldmath{$\sigma$}}}
\def\bdelta{\mbox{\boldmath{$\delta$}}}
\def\nn{\nu_{n^{\prime}}-\nu_n}
\def\mag{{\mbox{\Large{\it m}}}}
\def\dn{\delta {\mbox{{\Large {\it n}}}}}
\def\dm{\delta {\cal{M}}}
\def\bp{{\mbox{\boldmath{$\pi$}}}}
\begin{document}
\title{Ordered Phase in the Fermionized Heisenberg Antiferromagnet}

\author{S.~Azakov $^{1,2,}$\footnote{e-mail addres: azhep@lan.ab.az},
M.~Dilaver $^{2}$ , A.~M.~{\"O}zta{\c
s}
$^{2,}$\footnote{e-mail address: oztas@hacettepe.edu.tr}}
\maketitle

$^{1}${\it Institute of Physics, Azerbaijan Academy of Sciences, Baku,
Azerbaijan}\\

$^{2}${\it Hacettepe University, Physics Department, 06532, Beytepe,
Ankara, Turkey}\\ \\

\begin{abstract}
Thermal properties of the ordered phase of the spin $1/2$ isotropic Heisenberg 
Antiferromagnet on a d-dimensional hypercubical lattice  are studied
within the fermionic representation when the constraint of single
occupancy condition is taken into account by the method suggested by Popov
and Fedotov. Using saddle point approximation in path integral approach we
discuss not only the leading order but also the fluctuations around the
saddle point at one-loop level. The influence of taking into account the
single occupancy condition is discussed at all steps.
\end{abstract}

Quantized spin systems -75.50.Ee Antiferromagnetics \\ \\

\section{Introduction}
The two-dimensional spin-1/2 Heisenberg antiferromagnet (HAFM) on the
square lattice has been extensively studied during the last few years. The
motivation for this study stems from the discovery of high $T_c$
superconductivity in the ceramic compounds, where the
competition between superconductivity and antiferromagnetic order has been
observed experimentally \cite{Manous}.
\par
Contrary to early suggestions there is nowdays strong evidence that the
ground state of the fully isotropic quantum spin -1/2 HAFM on a two
dimensional regular lattice is the N\'eel state (the
classical
ground state (the N\'eel state) is not disordered by quantum
fluctuations.).
\par
This evidence is mainly based on numerical work
\cite{RY}. Recently it has
gained an additional support by results obtained analytically with the
help of various techniques e.g. large spin expansion, field theory of the
quantum nonlinear $\sigma$ -model \cite{CHN}, effective Hamiltonian
approach \cite{CTVV}, a modified mean
field approach (saddle point approximation) based on bosonic \cite{AA} or
fermionic representations \cite{AM} of spin operators.
\par
The main problem in the technique based on these
representations is to take into account the so-called single occupancy
condition.
\par
The aim of this work is to study the thermal properties of the ordered
(magnetic) phase of the spin $1/2$ isotropic HAFM on a $d$-dimensional
hypercubical  lattice \footnote{For simplicity we consider a simple 
hypercubical lattice though
our approach may be also used for a non-bipartite lattice} with periodic
boundary 
conditions 
within the fermionic representation when the constraint
of single occupancy condition is taken into account by the method
suggested by Popov and Fedotov \cite {PF}. We use saddle point
approximation and discuss not only the leading order but also the
fluctuations around the saddle point at the one-loop approximation
level. We show that at zero temperatures one-loop corrections to the
saddle point in our path integral description is equivalent to
next-to-leading order in the linear spin wave theory. 
At all steps we discuss the influence of taking into account the single
occupancy condition comparing the results of our calculations with those 
when this condition is disregarded. In particular we show that at finite
temperatures taking into account
the single occupancy condition considerably reduces the specific heat. 

For $T \neq 0$ the two-dimensional spin system has no long range order 
(the N\'eel state is destroyed by thermal fluctuations)\cite{MW} and its
state has to be treated as a paramagnetic one with strong
antiferromagnetic correlations at finite distances. So our finite
temperature results are relevant for the case when $d \geq 3$ and 
$T<T_N$, where $T_N$ is the N\'eel temperature.

\par 
In Sec.2 we briefly review the fermionization procedure of spin operators
by the method of Popov and Fedotov.
\par
In Sec.3 we discuss the mean field result (the leading order of the saddle
point approximation).
\par
In Sec.4 we obtain the one-loop corrections (Gaussian fluctuations) to
free energy and show that one can get the spin wave spectrum at zero
temperture. We also find the specific heat and discuss the influence of
the single occupancy condition on its temperature dependence.
\par
The last section is devoted to brief comments on our results.

\section{Fermionization by Popov and Fedotov' s method
and bosonic path integrals for the partition function.}

The Hamiltonian of the isotropic HAFM reads

\beeq
\hh_{s}=J\sum_{\la i,j \ra}\hbS_{i}.\hbS_{j} ,
\label{hs}
\eneq
the sum runs over ordered nearest neighbor sites of the d-dimensional
finite regular lattice with $M$ sites. For spin variables $\bS_{i} $ we
assume periodic boundary conditions, $ J > 0$ .

Many authors have proposed to use different representations of spin
operators by Bose or Fermi operators. However, the fact that the
dimensionality of the space in which these operators act is always greater
than the dimensionality of the space of spin operators leads to the
problem of the elimination of the superfluous states. Usually it is done
by putting some constraints on  the states.
In the present paper we choose fermionic representation of spin operators
\beeq
\hbS_{i}=\frac{1}{2}\hcd_{i \alpha}\bsigma_{\alpha \beta}\hc_{i
\beta},~~~~~~\alpha , \beta = 1(\uparrow) , 2(\downarrow) ,
\label{svec}
\eneq
the summation with respect to repeated Greek indices is assumed,
$\bsigma=(\sigma^{x},\sigma^{y},\sigma^{z})$ are the Pauli matrices, and
if we use this representation in the spin Hamiltonian  $\hh_{s}$ we shall
get the fermionic Hamiltonian:
\beeq
\hh_{F}=\frac{J}{4}\sum_{<i,j>}(\hcd_{i\alpha}\bsigma_{\alpha
\beta}\hc_{i\beta})(\hcd_{j\gamma}\bsigma_{\gamma
\delta}\hc_{j\delta})
\label{hf}
\eneq
$\hc_{i\alpha}$ and $\hcd_{i\alpha}$ are fermionic annihilation and
creation
operators (at site $i$ with spin projection $\alpha$ to the z axis), which
obey canonical anticommutation relation
$$
\{ \hc_{i\alpha},\hcd_{j\beta} \}=\delta_{i,j}\delta_{\alpha\beta} .
$$
Popov and Fedotov \cite{PF} proved that the partition function of the
model (1)
\begin{subequations}
\beeq
Z={\rm{Tr}}_{S}(e^{-\beta \hh_{S}})
\label{za}
\eneq
can also be written as
\beeq
Z=i^{M}{\rm{Tr}}_{F}(e^{-\beta \hh_{F}-i \frac{\pi}{2}\hat{N}}) .
\label{zb}
\eneq
\end{subequations}

In these formulas $\rm{Tr}_{S}(Tr_{F})$ is a trace in space where spin
(fermionic)
operators act,
$\hat{N}=\sum_{i=1}^{M}\hcd_{i\alpha} \hc_{i\alpha}
=\sum_{i=1}^{M} \hat{n}_{i}$ is the number operator.

Let us briefly repeat their arguments.
It is sufficient to consider only one site (we omit the site index). For spin
$\frac{1}{2}$, spin operators $\hat{S}^{a}(a=x,y,z)$ act in two
dimensional
space. But the space where fermionic operators act is four dimensional; we
have
states
$$\vert 0,0\ra,~~~\hcd_{\uparrow}\vert 0,0\ra=\vert \uparrow,0\ra,~~~
\hcd_{\downarrow}\vert 0,0\ra=\vert 0,\downarrow\ra,~~~
\hcd_{\uparrow}\hcd_{\downarrow}\vert 0,0\ra=\vert\uparrow ,\downarrow\ra 
 .$$

States $\vert \uparrow,0\ra, \vert 0,\downarrow\ra $ can be identified with
eigenstates of ${\hat{S}}^{z}$ operator with spin up and spin down , we
call
them $physical$  and denote $ \vert phys \ra$. Then states $\vert 0,0\ra$
and $\vert\uparrow ,\downarrow\ra $ are superfluous or $unphysical$ and
their contribution should be excluded.

The physical states span a two-dimensional physical subspace, characterized
by the single occupancy condition 
$$ \hat{n}\vert phys\ra =\vert phys\ra .$$

The direct product of the physical subspaces of all the sites form
the sectors in which the Hamiltonians $\hh_{S}$ and $\hh_{F}$ coincide.

In order to prove the basic formula Eq.(\ref{zb}) we write 
$$ \hh_{F}=\hh_{Fi}+\hh_{Fi}^{\prime}~~~,~~~\hat{N}=\hat{n}_{i}
+\hat{N}_{i}^{\prime} ,$$
where $\hh_{Fi}( \hat{n}_{i} )$ is that part of the $\hh_{F}(\hat{N})$
which
contains the fermionic operators of the i-th site and $\hh_{Fi}^{\prime}
(\hat{N}_{i}^{\prime})$ is the remaining part. For the Hamiltonian of HAFM
we have $$\hh_{Fi}\vert unphys \ra_{i}=0 .$$
Therefore, the trace in Eq.(\ref{zb}) taken over unphysical states of the
$i$-th site vanishes

$$ {\rm{Tr}}_{i~unphys} \{e^{- \beta \hh_{F}-i \frac{\pi}{2}{\hat{N}}} \}
={e^{- \beta \hh_{Fi}^{\prime}-i \frac{\pi}{2}{\hat{N_{i}^{\prime}}}}}
{\rm{Tr}}_{i~unphys} \{(-i)^{\hat{n}_{i}} \}=0 ,$$
since $ {\rm{Tr}}_{i~unphys} \{(-i)^{\hat{n}_{i}} \}=(-i)^{0} + (-i)^{2}=0
 .$

As a result , in the calculation of the trace all the unphysical
states are eliminated, while on the physical states $\hh_{F}=\hh_{S}$
and $\hat{N}\vert phys \ra=M\vert phys \ra$. Therefore

$${\rm{Tr}}_{F} ( e^{- \beta \hh_{F}-i \frac{\pi}{2}{\hat{N}}} 
)=(-i)^{M}{\rm{Tr}}_{phys}(e^{- \beta \hh_{F}} )=
\frac{1}{i^{M}}{\rm{Tr}}_{S}(e^{- \beta \hh_{S}})$$
which proves Eq.(\ref{zb}).

The evaluation of fermionic trace ${\rm{Tr}}_{F}$ requires only the
standard
technique  because this trace is unrestricted. It can be represented
as a path integral in terms of Grassmann fields $\eta$ and
$\overline{\eta}$ \cite{Azak}
\beeq
Z=i^M\int\mbox{$\cal{D}$} \mu_{ \eta} \exp {\left\{
-\int^{\beta}_{ 0}d\tau \left [ \sum_{i,\alpha}
\overline{\eta}_{i 
\alpha}( \tau)\left( \partial_{\tau}+i\frac{ \pi}{2 \beta}\right)
\eta_{i
\alpha}(\tau)+ \mbox{$\cal{H}$}_{F}(\overline{\eta},\eta;\tau) 
\right ] \right\} },
\label{zeta}
\eneq
where
\beqn
\mbox{$\cal{H}$}_{F}(\overline{\eta},\eta;\tau)&=&\frac{J}{4}\sum_{\la
i,j \ra}(\overline
{\eta}_{i\alpha}(\tau)\bsigma_{\alpha \beta}\eta_{i\beta}(\tau))
(\overline
{\eta}_{j\gamma}(\tau)\bsigma_{\gamma \delta}\eta_{j\delta}(\tau))
\cr&=&J\sum_{\la i,j \ra}\bS_{i}(\tau) \cdot \bS_{j}(\tau)
\label{hfeta}
\eeqn
and
\beeq
\bS_{i}(\tau) \equiv \frac{1}{2}\overline{\eta}_{i\alpha}(\tau)\bsigma_{\alpha
\beta}\eta_{i\beta}(\tau),
\label{sitau}
\eneq
$$\mbox{$\cal{D}$} \mu_{ \eta}=\prod_{0 \leq \tau \leq
\beta}~~~\prod_{i,\alpha} d \overline{\eta}_{i \alpha}(\tau) 
d \eta_{i \alpha}(\tau) .$$
Now let us do the Fourier transformation
\beeq
\bS_{i}(\tau)=\frac{1}{\beta} \SBZ \sum_{m} \bS(\qv,\om)e^{-i \om
\tau}e^{i
\qv. {\bf r}_{i}},
\label{sifou}
\eneq
where $\qv$ is the wave vector in  reciprocal space (we can restrict it to
the first Brillouin zone (BZ)), $\om=\frac{2 \pi m}{\beta}$ is a Matsubara
frequency for Bose field.

Summation over ordered nearest neighbors can be written as
\beeq
\sum_{\la i,j \ra}=\frac{1}{2} \sum_{i,\bdelta}~~ ,
\label{sum1}
\eneq
since $j$ is a nearest neighbor of $i$: $ {\bf r}_{j}={\bf r}_{i}+\bdelta
$, and $ \bdelta $  represents the displacement of $z=2d$ nearest
neighbors of each site . Then  
\beeq
\int^{\beta}_{0} d \tau \mbox{$\cal{H}$}_{F}(\overline{\eta},\eta;\tau)
=\frac{JMd}{\beta} \SBZ \sum_{m} \gqv\bS(\qv,\om)\bS(-\qv,-\om),
\label{inthf1}
\eneq
where the so-called structure function
$\gqv=\frac{1}{z}\sum_{\bdelta}e^{i\qv 
\bdelta}= \gamma_{-\qv}= \frac{1}{d}(\cos q_1+\cos q_2+\cdots+ \cos q_d).
$ 
From Eq.(\ref{sifou}) it follows that 
\beeq
\bS(-\qv,-\om)=\bS^{\star}(\qv,\om)
\label{sher}
\eneq
and if we write 
\beeq
\bS(\qv,\om)={\mbox{Re}}\bS(\qv,\om)+i{\mbox{Im}}\bS(\qv,\om)
\label{scomp}
\eneq
then
\beeq
\int^{\beta}_{0} d \tau \mbox{$\cal{H}$}_{F}( \overline{\eta},\eta;\tau)
=\frac{dJM}{\beta} \SBZ \sum_{m} \gqv [({\mbox{Re}}
\bS(\qv,\om))^{2}+({\mbox{Im}} 
\bS(\qv,\om))^{2}] .
\label{inthf2}
\eneq

\par
The standard way to decouple four fermion terms is to use
Hubbard-Stratonovich representations and introduce some auxiliary Bose
fields. The decoupling scheme is not unique and the particular choice of
the Bose fields depends which mean field solutions (ordered or disordered
for our model) we are going to discuss. Of course, before one starts to
use some approximation (usually saddle point approximation) all
representations are equivalent and if we are able to calculate the path
integrals exactly we shall get the same result.
\par
In the present paper we are considering the ordered phase (the disordered
phase which is the most relevant for $d\le 2$ will be discussed elsewhere
\cite{ADO2}) and  the Hubbard-Stratonovich decoupling can be done
with the help of an auxiliary vector field $ \mv\qom$ which plays the role
of the staggered magnetization
\beqn
e^{ - \int^{\beta}_{0} d \tau 
{\mbox{$\cal{H}$}}_{F}(\overline{\eta},\eta;\tau)} 
&=& 
\int {\mesm} \exp \left\{ \sum_{\qv,m} \frac{}{}\left[\frac{}{} -\vert \mv
\qom
\vert^{2}
\right.\right.  \nonumber \\
&+& \left.\left.  
  \sqrt{-\frac{dJM}{\beta} \gqv} [ \mv^{\star} {\qom} \bS {\qom}
+ \bS^{\star} 
{\qom} \mv {\qom} ] \right] \right\} ,  
\label{hubst}
\eeqn
and the path integration measure 
\beeq
{\mesm}={\prod_{ m}} { \prod_{ \qv \in BZ}}\quad {\prod_{a=x,y,z}}
\frac{d {\mbox{Re}} { \mbox{$\cal{M}$}}^{a}\qom d {\mbox{Im}} {
\mbox{$\cal{M}$}}^{a}\qom}{\pi} .
\label{mesm}
\eneq
For Fermi fields Fourier transformations are
\begin{subequations}
\beeq
\eta_{i \alpha}(\tau)=\sum_{n} \eta_{i \alpha}(\nu_{n})
e^{-i \nu_{n} \tau},
\label{eta1}
\eneq

\beeq
\overline{ \eta}_{i \alpha}( \tau)=\sum_{n}\overline{
\eta}_{i \alpha}(\nu_{n})e^{i \nu_{n} \tau},
\label{eta2}
\eneq
\end{subequations}
where $ \nu_{n}=\frac{2\pi(n+1)}{\beta}$ is a Matsubara frequency for 
Fermi fields. Then
\beeq
\bS\qom=\frac{\beta}{2M}\sum_{i,n^{\prime},n}\overline{\eta}_{i
\alpha}
(\nu_{n^{\prime}})\bsigma_{\alpha \beta} \eta_{i \beta}(\nu_{n})
e^{- i \qv {\bf r}_{i}}\delta_{\om,\nn}.
\label{sqm}
\eneq
and for the partition function we get

\beeq
Z= i^M \int {\cal{D}} \mu_{ \eta}{\mesm} \exp{ \left\{-
\sum_{\qv,m}{\vert \mv(\qv,\om ) \vert }^2 +  \sum_{i,j}
\sum_{n^{\prime},n}\overline{\eta}_{i \alpha}(\nu_{n^{\prime}})
{\cal{K}}_{n^{\prime},n;i,j}^{\alpha \beta}(\mv)\eta_{j \alpha}
(\nu_{n})\right\} } ,
\label{zmeta}
\eneq
where
\beeq
\mbox{$\cal{K}$}_{n^{\prime},n;i,j}^{\alpha \beta}(\mv)=\left[
\delta^{\alpha \beta} \left(-i \nu_{n} \beta +i \frac{\pi}{2}\right)
\delta_{n^{\prime}n}+\sum_{\qv}\sqrt{-\frac{dJ\beta}{M}\gqv}
e^{i \qv {\bf r}_{i}}\mv(\qv,\nn)\bsigma_{\alpha \beta} \right] 
\delta_{ij} ,
\label{km} 
\eneq
and the measure for the path integration with respect to Grassmann
variables now reads 
\beeq
\mbox{$\cal{D}$} \mu_{ \eta}=\prod_{i,n,\alpha} d \overline{\eta}_
{i \alpha}(\nu_{n}) d \eta_{i \alpha}(\nu_{n}).
\label{mesmnu}
\eneq
Integrating with respect to them we obtain
\beeq
Z= i^M\int {\mesm} e^{-S_{eff}[\mv]}=e^{-\beta F},
\label{zfree}
\eneq

where
\beeq
S_{eff}[ \mv]=\sum_{\qv,m}{ \vert \mv
\qom\vert}^2-{\rm Tr} \ln{\mbox{$\cal{K}(\mv)$}} 
\label{seffm}
\eneq
and $F$ is a free energy.

\section{The leading order of the saddle point approximation}

In order to deal with the AFM solution we shall choose a 
frequency independent solution along the z-axis ($\bp = \underbrace{(\pi , \pi
,...,\pi )}_{d})$
\beeq
\mv\qom=\hat{z}\sqrt{dM \beta J}\mag
\delta_{\qv,{\mbox{\boldmath{$\pi$}}}} .
\label{mqom}
\eneq
The real parameter $\mag$ is a staggered magnetization.
Then 
\beeq
(\mbox{$\cal{K}$}^{MF}(\mv))^{\alpha \beta}_{n^{\prime},n;i,j}=
\left [ \delta^{\alpha
\beta}\left(-i\nu_{n}\beta+i\frac{\pi}{2}\right)+(-1)^{i}dJ\beta 
\mag \sigma^{z}_{\alpha \beta}\right ] \delta_{n^{\prime}n}\delta_{ij} ,
\label{kmf}
\eneq
and $(-1)^i = (-1)^{({\bf{r}}_{i})_1 + \cdots + ({\bf{r}}_{i})_d}$.
So the free energy in the mean field leading order takes a form
\beeq
F^{MF}(\mag)=dJM\mag^2-\frac{1}{\beta}\sum_{m}\ln\left[1+e^{\beta E_{m}}
\right] - \frac{M}{\beta}\ln i ~,
\label{mffree1}
\eneq
where $\{E_{m}\}$ is the spectrum  of the mean field Hamiltonian
\beeq
\hh^{MF}=\sum_{j,\alpha}\left(\omega_{j, \alpha}\hcd_{j \alpha} \hc_{j
\alpha}+
i\frac{\pi}{2 \beta} \hcd_{j \alpha} \hc_{j \alpha}\right),
\label{hmf}
\eneq
$$\omega_{j1}=(-1)^{j}dJ \mag, $$
$$\omega_{j2}=-(-1)^{j}dJ \mag. $$
The summation with respect to eigenvalues can be done easily with the
result
\beeq
\sum_{m}\ln\left[1+e^{\beta E_{m}}\right]=M \ln{ \left( \frac{2}{i}
\cosh{(d\beta
J\mag)} \right) }.
\label{mats}
\eneq
So for the free energy in the leading
order we get 
\beeq
F^{MF}(\mag)=dJM\mag^2-\frac{M}{\beta} \ln{( \cosh{(d \beta J \mag) })}
- \frac{M}{\beta}\ln 2
\label{mfree2}
\eneq
Minimization of $F^{MF}(\mag)$ yields the mean field
staggered magnetization
equation 
\beeq
\mag= \frac{1}{2} \tanh{(d \beta J  \mag)} .
\label{mag}
\eneq
Exactly  the same result for magnetization one obtains in the mean field
approach to the $S = 1/2$ Heisenberg model with the Hamiltonian 
Eq.(\ref{hs}) working in terms of spin variables. 

If the single occupancy condition is disregarded instead of
Eqs.(\ref{mfree2}) and Eqs.(\ref{mag}) we get

\begin{equation}
F_{0}^{MF}(\mag_{0})=dJM\mag_{0}^2-\frac{2M}{\beta} \ln{\left(
\cosh{\left(\frac{d}{2} \beta J
\mag_{0}\right) }\right)}-\frac{2M}{\beta}\ln{2}
\label{fzeromf}
\end{equation}
and

\begin{equation}
\mag_{0}= \frac{1}{2} \tanh{\left(\frac{d}{2} \beta J  \mag_{0}\right)} .
\label{magsub0}
\end{equation}

\section{One-loop corrections}

Now we write

\beeq
\mv\qom = \hat{z}\sqrt{dM\beta J}\mag
\delta_{\qv,{\mbox{\boldmath{$\pi$}}}} + \dmv\qom~,
\label{matrixM}
\eneq
where $\dmv\qom$ are fluctuations of the magnetization around the mean-field 
value (the
leading order)$\mag $ satisfying Eq.(\ref{mag}).
Then

\beqn
{\cal{K}}_{n^{\prime},n;i,j}^{\alpha\beta}(\dmv) = \left[
\delta^{\alpha\beta}
\left(-i\nu_{n}\beta + i \frac{\pi}{2}\right) + 
(-1)^{j}dJ\beta\mag\sigma_{\alpha\beta}^{z}\right]
\delta_{n^{\prime}n}\delta_{ij}  
+ \dn_{ij}^{\alpha\beta}(\nn) ,
\label{kernelK}
\eeqn

where

\beeq
\dn_{ij}^{\alpha\beta}(\nn) \equiv \sum_{\qv \in
BZ}\sqrt{-\frac{dJ\beta}{M}\gqv} 
e^{i\qv\cdot{\rm{\bf{r}}}_i}\dmv{(\qv,\nn)}\bsigma_{\alpha\beta}
\delta_{ij} .
\label{deltan}
\eneq
The partition function

\beeq
Z = e^{-\beta F^{MF}}\int{\mesm}e^{-S_{eff}^{(2)}[ \dmv ]} ,
\label{partf}
\eneq

where

\beeq 
S_{eff}^{(2)} [ \dmv ] = \sum_{m}\SBZ | \dmv ( \qom )|^2 - {\rm{Tr}}\ln
{\cal{K}}
(
\dmv )~~,
\label{squantum}
\eneq

\beeq
{\mesm} \equiv \prod_{m}\prod_{\qv \in BZ} \frac{d {\mbox{Re}}\dmv \qom  d
{\mbox{Im}}\dmv \qom}{\pi}~~~.
\label{measure}
\eneq
The superscript $(2)$ means that only the terms of the second order with
respect to $\dmv$ are kept. Thus we take into account only so-called
Gaussian fluctuations.

Let us define a matrix $G$ such that its matrix elements has a form

\beeq
G_{n^{\prime},n;i,j}^{\alpha\beta} = \left[-i \nu_{n}\beta + i
\frac{\pi}{2}
- (-1)^{\alpha}(-1)^{j}dJ\beta\mag\right]^{-1}
\delta_{ij}\delta_{\alpha\beta}\delta_{n^{\prime}n} .
\label{gmatrix}
\eneq
This matrix is the one particle propagator evaluated at the saddle point.
Then Eq.(\ref{kernelK}) written in the matrix form (with respect to spin
index
$\alpha$ and lattice site index $i$ ) takes a form 

\beeq
{\cal{K}}_{n^{\prime}n} = G_{n^{\prime}n}^{-1} - \dn ( \nn )
\label{knpn}
\eneq
and

\beqn
&&{\rm{Tr}}\ln {\cal{K}}({\dmv}) = {\rm{Tr}}\ln G^{-1} + {\rm{Tr}}\ln(1 -
G\dn )
\nonumber \\
&& = {\rm{Tr}}\ln G^{-1} - {\rm{Tr}} [G\dn ] - \frac{1}{2} {\rm{Tr}} [ G
\dn G \dn ] -
\cdots \quad .
\nonumber
\label{tracek}
\eeqn
 
The term which describes the Gaussian fluctuations in more explicit form
reads

\beeq
{\rm{Tr}} [ G \dn G \dn ] = \sum_{n,m}{\rm{Sp}} [(\overline{\mag} - i
\tilde{\nu}_{n}\beta
)^{-1} \dn (\om ) (\overline{\mag} - i \tilde{\nu}_{n}\beta + i \om \beta
)^{-1}
\dn (- \om ) ] ,
\label{gdngdn}
\eneq

where ${\overline{\mag}}$ is a matrix in the space of $i$ and $\alpha$
indices
with the matrix element

\beeq 
{\overline{\mag}}_{ij}^{\alpha\beta} = - (-1)^{\alpha}(-1)^{j} d J \beta
\mag \delta_{ij} \delta_{\alpha\beta} ,
\label{mbar}
\eneq

and Sp is a trace in this space ( its element we denote as $|i\alpha
\ra$),

$$ {\tilde{\nu}}_{n} \equiv \nu_{n} + \frac{\pi}{2\beta} .$$ 

So in the one-loop approximation we have for $S_{eff}^{(2)}[\dmv ]$

\beeq
S_{eff}^{(2)} = \sum_{m,\qv \in BZ} | \dmv (\qom ) |^2 +
{\tilde{S}}_{eff}^{(2)} [\dmv ]
\label{eqseff1}
\eneq

and
\beeq
{\tilde{S}}_{eff}^{(2)}[\dmv ] = \frac{1}{2}\sum_{i,\alpha , \beta , m}
T_{\alpha\beta}(i,m)\langle i\alpha | \dn ( \om ) | i\beta \rangle \langle
i\beta | \dn ( -\om ) | i\alpha \rangle ,
\label{tildeS}
\eneq

where

\beqn
T_{\alpha\beta}(i,m) &\equiv& \sum_{n} \langle i\alpha | (\overline{\mag}
-
i {\tilde{\nu}}_{n}\beta )^{-1} | i\alpha \rangle
\langle i\beta | (\overline{\mag} -
i {\tilde{\nu}}_{n}\beta + i \om \beta )^{-1} | i\beta \rangle 
\nonumber \\
&=& \sum_{n}\frac{1}{i\tilde{\nu}_{n}\beta +
(-1)^{\alpha}(-1)^{i}dJ\beta\mag}~\frac{1}{i\tilde{\nu}_{n}\beta
- i\om\beta + (-1)^{\beta}(-1)^{i}dJ\beta\mag}~.
\label{tabeta}
\eeqn

The summation with respect to Matsubara frequencies can be easily done
with the following result

\beqn
\Phi(A,B;\om ) &\equiv& \sum_{n}\frac{1}{i{\tilde{\nu}}_n \beta - A}
~\frac{1}{i{\tilde{\nu}}_n \beta -B - i\om\beta} \nonumber \\
&=&\frac{1}{i\om\beta - A +
B}~\frac{\sinh{[\frac{1}{2}(A-B)]}}{i\sinh{[\frac{1}{2}(A+B)]} +
\cosh{[\frac{1}{2}(A-B)]}} , \nonumber
\eeqn

if $A \neq B$, and

\beeq
\Phi(A,A;\om ) = -\delta_{m0}\frac{1 - i\sinh{A}}{2\cosh^2{A}} .
\eneq

We need to know only the expression for the special choice: $B = -A$. In
this
case

\beeq
\Phi (A, -A; \om ) = - \frac{2A\tanh{A}}{(\om\beta )^2 + 4 A^2} -
\frac{i\om \beta \tanh{A}}{(\om\beta )^2 + 4 A^2} .
\label{phieqn}
\eneq

So we get $( A \equiv dJ\beta\mag )$
\begin{subequations}
\beeq
T_{11}(j,m) = -\delta_{m0}\frac{1 - i (-1)^{j}\sinh{A}}{2\cosh^2 A} \equiv
-\delta_{m0}[\kappa - i (-1)^{j}\rho ] ,
\label{t11}
\eneq

\beeq
T_{22}(j,m) = -\delta_{m0}[\kappa + i (-1)^{j}\rho ] ,
\label{t22}
\eneq       
\label{tdiag}
\end{subequations}
\begin{subequations}
\beeq
T_{12}(j,m) = - \frac{2A\tanh{A}}{( \om \beta )^2 + 4 A^2} - \frac{i
(-1)^j \om \beta \tanh{A}}{( \om \beta )^2 + 4 A^2} \equiv
\xi{(m)} + i
(-1)^j \zeta{(m)}
\label{t12}
\eneq
and
\beeq
T_{21}(j,m)= \xi{(m)} - i (-1)^j \zeta{(m)} .
\label{t21}
\eneq
\label{tcross}
\end{subequations}
Taking into account the Eq.(\ref{mag}) for the mean field magnetization we
get

$$ \kappa = \frac{1}{2}( 1 -4 \mag^2 )~. $$
If the single occupancy condition is 
neglected we get instead 
($A_0 \equiv dJ\beta\mag_0$ )

$$\kappa_0 = \frac{1}{4\cosh^2 \frac{A_0}{2}} = \frac{1}{4}( 1 - 4 \mag_0^2 )$$

and

$$\xi_0 (m) = - \frac{2A_0 \tanh\frac{A_0}{2}}{( \om\beta )^2 + 4
A_0^2}~~,
~~ \zeta_0 (m) = - \frac{\om\beta\tanh\frac{A_0}{2}}{( \om\beta )^2 + 
4 A_0^2}~.$$		

From Eqs.(\ref{mag}) and (\ref{magsub0}) it follows that $\xi(m)|_{A=A_0} 
= \xi_{0}(m)$ and
$\zeta(m)|_{A=A_0} = \zeta_{0} (m) $.

From Eq.(\ref{deltan}) we have

\begin{subequations}
\beeq
\langle i1 | \dn ( \om ) | i1 \rangle = \sum_{\qv \in BZ}\sqrt{-
\frac{dJ\beta}{M}\gqv}e^{i\qv{\rm{\bf{r}}}_i}{\dm}^{z} \qom , 
\label{i1i1}
\eneq

\beeq
\langle i2 | \dn ( \om ) | i2 \rangle = - \sum_{\qv \in BZ}\sqrt{-
\frac{dJ\beta}{M}\gqv}e^{i\qv{\rm{\bf{r}}}_i}\dm^{z} \qom ,
\label{i2i2}
\eneq            
\label{idiag}
\end{subequations}

\begin{subequations}
\beeq
\langle i1 | \dn ( \om ) | i2 \rangle = \sum_{\qv \in BZ}\sqrt{-
\frac{dJ\beta}{M}\gqv}e^{i\qv{\rm{\bf{r}}}_i}[\dm^{x} \qom  - i \dm^{y}
\qom  ] ,
\label{i1i2}
\eneq

\beeq
\langle i2 | \dn ( \om ) | i1 \rangle = \sum_{\qv \in BZ}\sqrt{-
\frac{dJ\beta}{M}\gqv}e^{i\qv{\rm{\bf{r}}}_i}[\dm^{x} \qom  + i \dm^{y}
\qom  ]~,                                                                   
\label{i2i1}
\eneq
\label{icross}
\end{subequations}       
and $\tilde{S}_{eff}^{(2)}$ defined in Eq.(\ref{tildeS}) can be rewritten as

\beqn
\tilde{S}_{eff}^{(2)} & = & \tilde{S}_{L}^{(2)} + \tilde{S}_{T}^{(2)}~,
\nonumber 
\eeqn
where the longitudinal part
\beqn
\tilde{S}_{L}^{(2)} & = & \frac{1}{2}\sum_{i,m}\left\{T_{11}(m)\langle i1
| \dn
(\om) | i1 \rangle \langle i1 | \dn (-\om) | i1 \rangle
+ T_{22}(m)\langle i2 | \dn
(\om) | i2 \rangle \langle i2 | \dn (-\om) | i2 \rangle \right\} \nonumber
\eeqn
and the transverse part
\beqn
\tilde{S}_{T}^{(2)} & = & \frac{1}{2}\sum_{i,m}\left\{T_{12}(m)\langle i1
| \dn
(\om) | i2 \rangle \langle i2 | \dn (-\om) | i1 \rangle
+ T_{21}(m)\langle i2 | \dn
(\om) | i1 \rangle \langle i1 | \dn (-\om) | i2 \rangle \right\} ~~. \nonumber
\label{seff2}
\eeqn

So using Eqs.(\ref{tdiag}) and (\ref{idiag}) we obtain

\beeq
\tilde{S}_{L}^{(2)} = dJ\beta\kappa\sum_{\qv \in BZ}\gqv | \dm^{z}( \qv ,
0 ) |^2
\label{sofl2}
\eneq
and with the help of Eqs.(\ref{tcross}) and (\ref{icross})

\beqn
\tilde{S}_{T}^{(2)} &=& \frac{1}{2}( -dJ\beta )\sum_{m,\qv \in BZ}\{\gqv
[\dm^{x}( \qom
)
- i \dm^{y}( \qom ) ] [\dm^{x}(-\qv,-\om ) + i \dm^{y}(-\qv,-\om )]\xi (m)
\nonumber \\
&& + \gqv [\dm^{x}( \qom ) + i \dm^{y}( \qom )][\dm^{x}(-\qv,-\om ) - i
\dm^{y}(-\qv,-\om )] \xi (m) \nonumber \\
&& - \gqv [\dm^{x}( \qom ) - i \dm^{y}( \qom )][\dm^{x}(-\qv -
{\mbox{\boldmath{$\pi$}}} ,-\om )
+ i \dm^{y}(-\qv -{\mbox{\boldmath{$\pi$}}} ,-\om )]\zeta (m) \nonumber \\
&& - \gqv [\dm^{x}( \qom ) + i \dm^{y}( \qom )][\dm^{x}(-\qv -
{\mbox{\boldmath{$\pi$}}} ,-\om )
- i \dm^{y}(-\qv -{\mbox{\boldmath{$\pi$}}} ,-\om )]\zeta (m)\}~~.
\nonumber    
\label{soft2}
\eeqn

So

\beqn
\tilde{S}_{eff}^{(2)} &=& dJ\beta\kappa\sum_{\qv \in RBZ} \vert \gqv \vert
\left\{ \vert \dm^{z}(\qv,0)\vert^2 - \vert \dm^{z}(\qv+\bp,0)\vert^2
\right\} \nonumber \\
&+& dJ\beta\sum_{m}\sum_{\qv \in RBZ} \vert \gqv \vert \left\{\frac{}{}
\left[
- \vert \dm^{x} \qom \vert^2 + \vert \dm^{x}(\qv+\bp,\om)\vert^2  
\right. \right. \\
&-& \left. \vert \dm^{y} \qom \vert^2 + \vert
\dm^{y}(\qv+\bp,\om)\vert^2 \right]\xi (m)  \nonumber \\
&+& 2{\rm{Im}}\left[\dm^{x} \qom \dm^{y} (-\qv-\bp,-\om) \right. \nonumber
\\
&-& \left.\left. \dm^{y} \qom \dm^{x}
(-\qv-\bp,-\om)\right]\zeta{(m)}\frac{}{}\right\}~~. \nonumber
\label{totseff}
\eeqn

The part of $S_{eff}^{(2)}$ which describes the transverse fluctuations
for each
$\qv$ vector and for each value of $m$ consists of four $2\times2$ blocks mixing the
real and imaginary components of $\dm^{x}$ and $\dm^{y}$ at $\qv$ and $\qv+\bp$ in
pairs. The matrices corresponding to these blocks are given by $(\qv \in
RBZ )$

\begin{equation}
\left(\begin{array}{cc}
1 - d\beta J \vert \gqv \vert \xi (m) & \pm d\beta J \vert \gqv \vert
\zeta (m)\\
& \\
\pm d\beta J \vert \gqv \vert \zeta (m) & 1 + d\beta J \vert \gqv \vert
\xi (m)
\end{array}\right)
\label{bmatrix}
\end{equation}
with the eigenvalues 

\beeq
\lambda_{\pm}(\qv,m) = 1 \pm \frac{2d\mag\vert \gqv \vert}{\sqrt{(\om
J^{-1})^2 +
(2d\mag)^2}}~.
\label{evallam}
\eneq
We see that $\lambda_{-}(\qv,m)$ vanishes at $\om = 0 $ when $\qv = 0$ .
The corresponding eigenmodes are the Goldstone modes (spin waves) which
appear due to the fact that AFM (ordered) phase is a phase with
spontaneously broken symmetry.

So from Eq.(\ref{partf}) we obtain up to some inessential constant the free 
energy including one-loop corrections

\beqn
F &=& F^{MF} + \frac{2}{\beta}\sum_{\qv \in RBZ} \sum_{m>0} \ln
(\lambda_{+}(\qv,m)\lambda_{-}(\qv,m)) + \frac{1}{\beta}\sum_{\qv
\in RBZ}'\ln (1- \gqv^2) 
\nonumber \\ 
&& + \frac{1}{2\beta}\sum_{\qv \in RBZ}\ln (1 - d^2\beta^2 J^2 \kappa^2
 \gqv ^2 )~,
\label{fenergy}
\eeqn
where $\sum_{\qv \in RBZ}'$ means that the point $\qv=0$ should be
excluded.
The summation with respect to Matsubara frequencies  can be done with help 
of the formula
\beeq
\sum_{m > 0}\ln\left(1 - \frac{A^2}{(\om\beta)^2 + B^2}\right) =
\ln \left( \frac{B\sinh\frac{\sqrt{B^2 - A^2}}{2}}{\sqrt{B^2 - A^2}\sinh
\frac{B}{2}}\right)
\label{matsubara}
\eneq
and we get finally for the free energy
\beqn 
F &=& F^{MF} + \frac{2}{\beta}\sum_{\qv \in RBZ}^{\prime} 
\ln \left[\frac{\sinh \left(d\mag\beta J \sqrt{1 -  \gqv^2}\right)}{\sinh
(d\mag\beta J)} \right] + \frac{2}{\beta} \ln \frac{d\mag\beta J}{\sinh
(d\mag\beta J)} 
\nonumber \\
&& + \frac{1}{2\beta}\sum_{\qv \in RBZ}\ln \left[ 1 - \frac{d^2}{4}\beta^2
J^2 ( 1 - 4\mag^2
)^2  \gqv^2 \right]~~.
\label{lastfree}
\eeqn

Now taking the limit of zero temperature $\beta \rightarrow \infty$ we get 
the energy of the ground state per site (there is no contribution to the
limiting expression from the last term)

\beeq
\frac{F}{M} \begin{array}{c} \\{\overrightarrow{T\rightarrow 0}}\end{array}
\frac{E_0^{(0)}}{M} = -
\frac{dJ}{4}
+ \frac{dJ}{M} \sum_{\qv \in RBZ} \left[ \sqrt{1 - \gqv^2} - 1
\right]~~,
\label{fperm}
\eneq
which is exactly the ground state energy per site obtained in linear spin-wave
approximation \cite{SW} for spin $1/2$.
\par 
It is easy to check that the same zero temperature result will be obtained
when the single occupancy condition is disregarded. 
\par
On the contrary, at finite temperatures taking into account the single
occupancy condition gives different values of the different
thermodynamical quantities, e.g. for the free energy this affects in
changing the temperature dependence of the magnetization $\mag$ and
changing the longitudinal part. Figs.1 and 2 show the difference of the
results for the cases when single occupancy condition is taken into
account (solid line) and when it is disregarded (dashed line). On Fig.1
the temperature dependence of the internal  energy and entropy and  on
Fig.2 of the specific heat are given. From our numerical calculations we
found that only in the interval $0\leq t \leq 0.13$ the difference is
negligebly small. In this interval the  specific heat goes to zero as $C_v 
= at^{\alpha}$ with $\alpha =3$, as it should \cite {SW}.
\section{Conclusion}

We have studied the magnetic (ordered) phase of the isotropic spin 1/2
HAFM defined on the simple d-dimensional hypercubic lattice using
fermionized spin operators and saddle point approximation. Single
occupancy condition which is needed when spin operators are bosonized or
fermionized is taken into account by the method of Popov and Fedotov.
\par
It is shown that inclusion of the one-loop corrections to the leading
order of the saddle point approximation leads in the limit of zero
temperature exactly to the same expression for the ground state energy
which
one obtains for the next-to-leading term in the linear spin wave theory,
and this result does not depend if the single occupancy condition is
disregarded or not.
\par
It is worthwile to mention that in the mean field theory of the spin 1/2
HAFM where Schwinger bosons are used \cite{AA} one obtains the same result
of the linear spin wave theory at zero temperature already in the leading
order and taking into account the single occupancy condition is crucial in
this case.
\par
We demonstrated that in our approach taking into account the single
occupancy condition changes finite temperature results 
considerably.\\

{\bf Acknowledgments}\\ 

S.A. is pleased to thank Y.G{\"u}nd{\"u}{\c c} for his kind invitation to
visit 
the Department of Physics Engineering of Hacettepe University and
providing a very stimulating atmosphere. Support for S.A. from the
Scientific and Technical Research Council of Turkey (T{\"U}B{\.I}TAK)
within
the framework of the NATO-CP Advanced Fellowship Programme is also
gratefully acknowledged. \\ \\

\leftline{\bf References}

\renewenvironment{thebibliography}[1]
	{\begin{list}{\arabic{enumi}.}
	{\usecounter{enumi}\setlength{\parsep}{0pt}
	 \setlength{\itemsep}{0pt}
         \settowidth
 {\labelwidth}{#1.}\sloppy}}{\end{list}}

\def\jnl#1#2#3#4{{#1}{\bf #2}, #3 (#4)}
\def\em{\it}
\def\nc{\em Nuovo Cimento }
\def\jpA{{\em J.\ Phys.} A}
\def\jpC{{\em J.\ Phys.\ Cond.\ Mat.} }
\def\npB{{\em Nucl.\ Phys.} B}
\def\plA{{\em Phys.\ Lett.} A}
\def\plB{{\em Phys.\ Lett.} B}
\def\prl{\em Phys.\ Rev.\ Lett. }
\def\pr{{\em Phys.\ Rev.} }
\def\prB{{\em Phys.\ Rev.} B}
\def\prD{{\em Phys.\ Rev.} D}
\def\ap{{\em Ann.\ Phys.\ (N.Y.)} }
\def\pps{{\em Proc.\ Phys.\ Soc.}}
\def\rmp{{\em Rev.\ Mod.\ Phys.} }
\def\zpC{{\em Z.\ Phys.} C}
\def\sci{\em Science}
\def\cmp{\em Comm.\ Math.\ Phys. }
\def\mplA{{\em Mod.\ Phys.\ Lett.} A}
\def\mplB{{\em Mod.\ Phys.\ Lett.} B}
\def\ijmpB{{\em Int.\ J.\ Mod.\ Phys.} B}
\def\IJMPB{{\em Int.\ J.\ Mod.\ Phys.} B}
\def\ijmpA{{\em Int.\ J.\ Mod.\ Phys.} A}
\def\IJMPA{{\em Int.\ J.\ Mod.\ Phys.} A}
\def\ptp{{\em Prog.\ Theoret.\ Phys.} } 
\def\Zphys{{\em Z.\ Phys.} }
\def\jpsJ{{\em J.\ Phys.\ Soc.\ Japan }}
\def\jmp{{\em J.\ Mod.\ Phys.} }
\def\jssc{{\em J.\ Solid State Chem.\ }}
\def\jMP{{\em J.\ Math.\ Phys.}}
\def\etal{{\em et al,} }

\vfill

\newpage
{\bf Figure captions}\\
Fig.1~~Internal energy per site $E/M$ and entropy per site $S/M$ versus
dimensionless temperature $t=(\beta J)^{-1}$ for the 3-dimensional cubic
lattice. Dashed and solid lines correspond to the cases when single
occupancy condition is disregarded and taken into account respectively.\\ 
\\
Fig.2~~ Temperature dependence of the specific heat per site $C_v/(K_Bt)$ 
for the 3-dimensional cubic lattice. Dashed and solid lines
correspond to the cases when single occupancy condition is disregarded
and taken into account respectively.\\
 
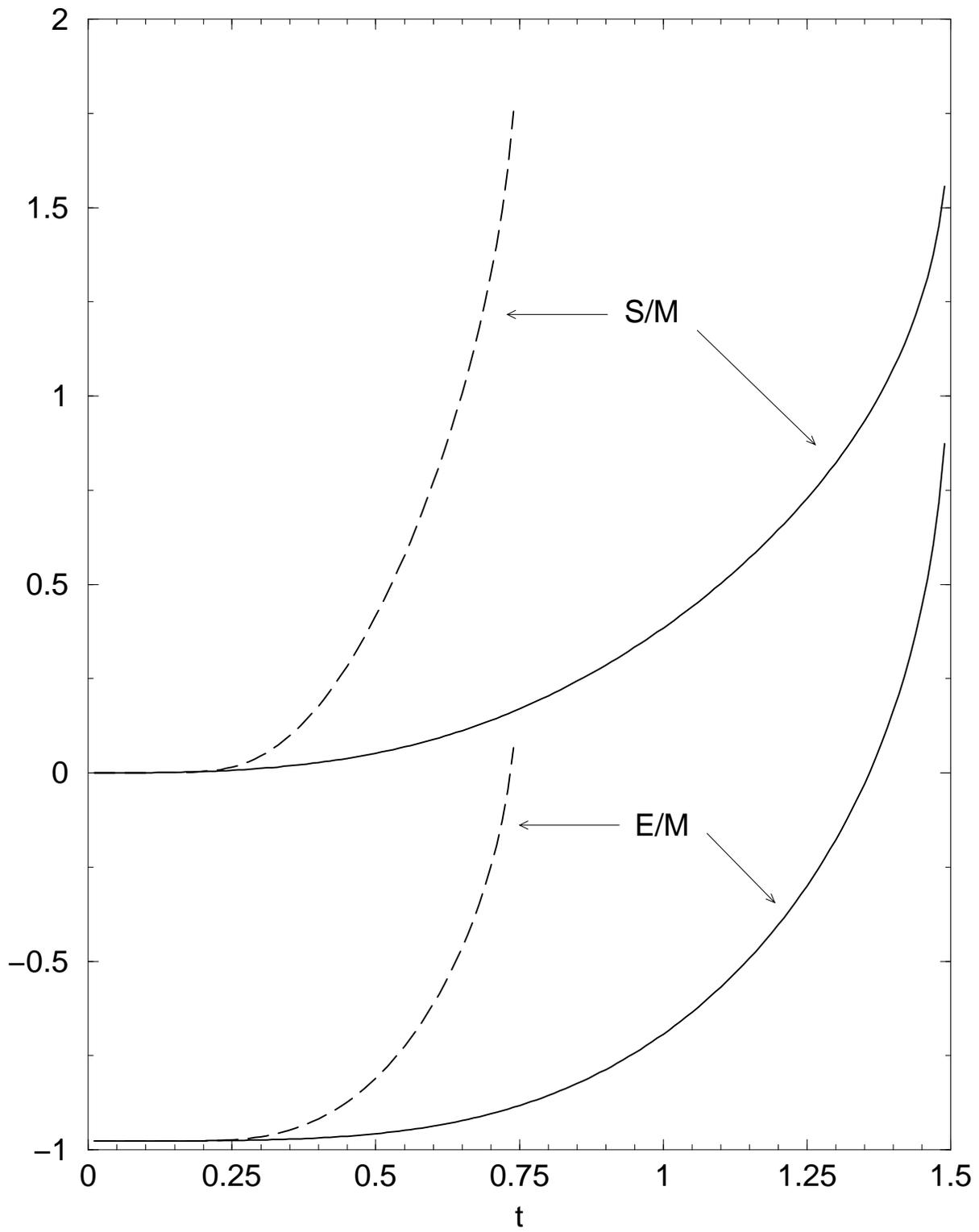
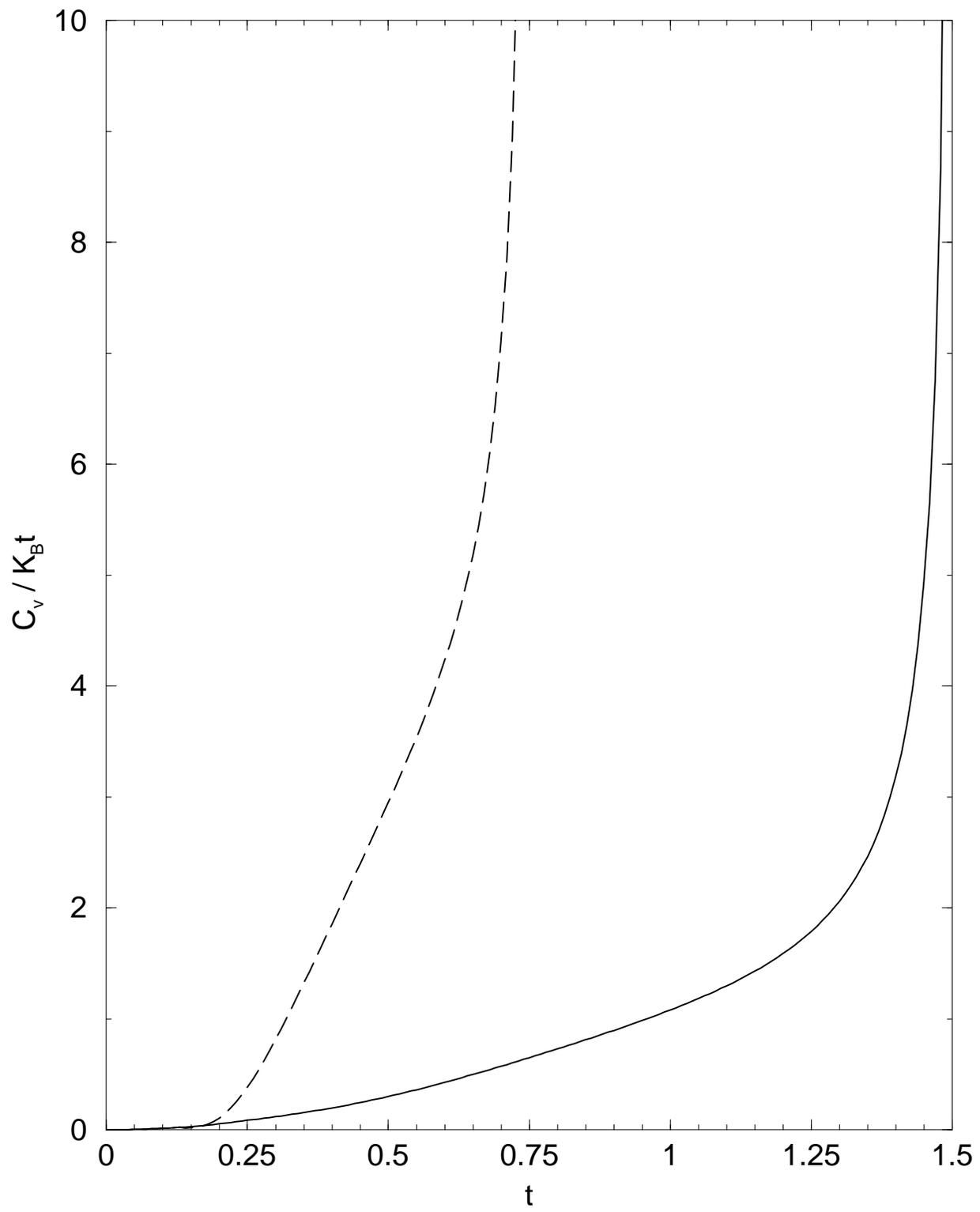
\end{document}